\documentclass[10pt,twocolumn]{article}
\usepackage{authblk}
\usepackage{amsmath}

\usepackage[export]{adjustbox}
\usepackage{graphicx}
\usepackage{color}
\usepackage{xcolor}
\usepackage{wasysym}
\usepackage{dcolumn}
\usepackage{bm}
\usepackage{hyperref}
\definecolor{mygreen}{rgb}{0,0.5,0}
\definecolor{mybrown}{rgb}{0.65,0.16,0.16}

\title{Average turbulence dynamics from a one-parameter kinetic theory}

\author{Hudong Chen}
\affil{Dassault Systemes Simulia Corp, 175 Wyman Street, Waltham, MA 02451, USA}

\author{Ilya Staroselsky}
\affil{Dassault Systemes Simulia Corp, 175 Wyman Street, Waltham, MA 02451, USA}

\author{Katepalli R. Sreenivasan}
\affil{Department of Mechanical and Aerospace Engineering, New York University, New York, NY, $11201$, USA}
\affil{Department of Physics and the Courant Institute of Mathematical Sciences, New York University, New York, NY $11201$, USA \\ krs3@nyu.edu}

\author{V. Yakhot}
\affil{Department of Mechanical and Aerospace Engineering, New York University, New York, NY, $11201$, USA}
\affil{Department of Mechanical Engineering, Boston University, Boston, MA 02215}

\begin{document}

\maketitle

\vspace{0.3in} 

\begin{abstract}
We show theoretically that the mean turbulent dynamics can be described by a kinetic theory representation with a single free relaxation time that depends on space and time. A proper kinetic equation is constructed from averaging the Klimontovich-type equation for fluid elements satisfying the Navier-Stokes hydrodynamics exactly. The turbulent kinetic energy plays the role of temperature in standard molecular thermodynamics. We show that the dynamics of turbulent fluctuations resembles a collision process that asymptotically drives the mean distribution towards a Gaussian (Maxwell-Boltzmann) equilibrium form. Non-Gaussianity arises directly from non-equilibrium shear effects. The present framework overcomes the bane of most conventional turbulence models and theoretical frameworks arising from the lack of scale separation between the mean and fluctuating scales of the Navier-Stokes equation with an eddy viscous term. An averaged turbulent flow in the present framework behaves more like a flow of finite Knudsen number with finite relaxation time, and is thus more suitably described in a kinetic theory representation.   


\end{abstract}
\vspace{0.3in} 

\section{Introduction}

Turbulent flows differ from laminar flows fundamentally via the appearance of large fluctuations on scales much larger 
than the microscale thermal fluctuations \cite {sain}. The effect of turbulent fluctuations on the mean flow has been 
modeled as an eddy viscosity since Boussinesq \cite{bous} and Reynolds \cite{reyn}, based on the analogy to thermal 
fluctuations of molecules in a gas; see also Prandtl \cite{pran}. These eddy viscosity ideas have been a pillar of 
theoretical and engineering models of turbulence to this day. It has, however, been clear for some time that eddy 
viscosity is largely a metaphor, and that more faithful physics is required for proper turbulence modeling. One 
hope had been that the limitations of eddy viscosity could be circumvented by starting with the Boltzmann equation 
\cite{bolt}, but that approach has been unsuccessful for reasons to be described shortly \cite{chen}. Instead, by 
starting with the Klimontovich-type kinetic equation \cite{klim}, whose average is the exact Navier-Stokes equation, 
we obtain a BGK-type relaxation equation for one-point probability density function (pdf) of turbulent fluctuation. 
This is the major contribution of this paper. The finite relaxation time is the only unknown parameter in this framework. 

As noted already, essentially all conventional turbulence models describing the mean turbulent flow dynamics have been 
based on the Navier-Stokes equation with an eddy viscosity together with the regular molecular viscosity. Specifically, 
one writes
\begin{equation}
\partial_t {\bf U} + {\bf U}\cdot \nabla {\bf U} 
= - \nabla {\bar p} + \nu \nabla^2 {\bf U}
+ \nabla \cdot (\nu_{eddy} {\bar {\bf S}} ),
\label{RANS0}
\end{equation}
where $\nu$ and $\nu_{eddy}$ are the molecular and the eddy viscosities, respectively. The latter is a consequence of 
modeling the (deviatoric part of) the Reynolds stress term of turbulent fluctuations by an eddy viscosity term according 
to the Boussinesq approximation
\begin{equation}
\langle \delta {\bf u}\delta {\bf u}\rangle \approx - \nu_{eddy} {\bar {\bf S}},
\label{bous}
\end{equation}
where $\langle \;\;\rangle$ denotes an ensemble average. In Eqs. ~(\ref{RANS0}) and (\ref{bous}), ${\bf U}$ denotes 
the average flow velocity, $\delta {\bf u}$ the fluctuation, and ${\bar {\bf S}}$ the rate-of-strain tensor of 
the mean flow. 

It is well recognized that eddy viscosity hypothesis requires a separation of scales between the mean and 
fluctuating fields. But simple estimates \cite{wilc} reveal that the effective mean free path of eddies 
(e.g., Prandtl's mixing length \cite{pran}) is comparable to the characteristic scale of the mean flow. 
Nevertheless, because of its simplicity, eddy viscosity remains the foundation of most models of the 
Reynolds averaged Navier-Stokes (RANS) type as well as large eddy simulations (LES) \cite{lesi}. 

There have been extensions beyond the framework of eddy viscosity to nonlinear models 
\cite{spez,yosh, rubi,yakh}, but the level of empiricism increases substantially. Alternatively, it may appear 
at first sight that the lack of separation of scales in turbulence can be accounted for more naturally by 
seeking a description in terms of the kinetic theory. 

Indeed, the mean dynamics generated by averaging the Boltzmann equation has flow features with finite 
Knudsen number (a measure of the ratio of the mean free path of molecules to the characteristic flow length 
scale). Using an expansion in small Knudsen number \cite{chap} one can typically obtain non-Newtonian terms, 
whose transport coefficients agree with the above nonlinear turbulence models. In fact, one can formally 
generate equations for the turbulent velocity as well as turbulent kinetic energy \cite{dego,chen,chen2}, 
and match the conditions to some known turbulent eddy viscosity model. But this approach has a difficulty 
of principle. If we agree that turbulence is a property of the Navier-Stokes equation, the averaged equations 
arising from the Boltzmann equation are "imposed" instead of being a description from first principles of 
turbulence. 

Another way of deriving continuum equations from the Boltzmann equation is to start from the regular 
Boltzmann equation for molecules. Such an approach was first attempted over two decades ago by 
successively averaging the regular Boltzmann equation to remove small scales \cite{chen3}. Unfortunately, 
this attempt was unsuccessful: Among other problems, the main issue is that the resulting averaged equilibrium 
distribution contains the temperature in the same place as the turbulent kinetic energy. Hence, the thermal 
molecular effects cannot be removed via some successive averaging procedure. The fundamental reason for 
this bottleneck is that molecules move approximately with the speed of sound ($\sim \sqrt{T}$), so that 
no naive averaging procedure can pick out the tiny hydrodynamic velocity from a microscopic velocity 
distribution of molecules, which is essentially isotropic with a standard deviation of $\sqrt{T}$. 
As a consequence, the resulting eddy viscosity is erroneously dependent on (thermal) temperature as 
opposed to turbulent hydrodynamic properties alone. The existence of dimensional quantities other than 
molecular viscosity, outside of the intrinsic turbulent flow properties, is also inconsistent with turbulent 
scaling properties in the limit of infinite Reynolds number (vanishing molecular viscosity). Therefore, the 
regular Boltzmann equation is at least an inconvenient starting point, if not an inappropriate one; or, there 
is a need of a fundamentally different averaging process.

In the current study, we take an alternative starting point. Instead of averaging the Boltzmann equation for molecules, we start from the framework of a micro-kinetic equation known as the Klimontovich equation \cite{klim}.  For a turbulent flow that comprises an infinite number of fluid elements of infinitesimal sizes, the Klimontovich equation describes the evolution of the pdf of such fluid elements. This equation is exactly the (unaveraged) Navier-Stokes equation. As we will show below, averaging the Klimontovich equation leads to a desired kinetic equation, without suffering from the mixing of dynamic scales with the micro thermal temperature. Equally important, the resulting kinetic equation provides a new framework for understanding the nature of turbulent fluctuations and turbulent eddies. This new formulation also puts the long-standing and heuristic analogy of turbulent eddies with molecules on better theoretical grounds and offers deeper insights into the nature of turbulent fluctuations and their effects on the averaged turbulent hydrodynamics. 

The formulation described in Sections II-V is followed by a discussion of its implications in Section VI.

\section{Basic formulation}

For a particle that moves exactly according to the Navier-Stokes equation, in the description of Klimontovich, the pdf of particles, $f = f({\bf x}, {\bf v}, t)$, obeys the equation of motion given by
\begin{equation}
\partial_t f + {\bf v}\cdot\nabla f + {\bf a}\cdot\nabla_{\bf v} f = 0.
\label{klim}
\end{equation}
If the fluid flow is self-consistently generated by the motion of such particles, we have
\begin{equation}
\int d{\bf v} f({\bf x}, {\bf v}, t) = 1,
\label{rho}
\end{equation}
corresponding to the normalization of the pdf for an incompressible flow. We also have
\begin{equation}
\int d{\bf v} {\bf v} f({\bf x}, {\bf v}, t) = {\bf u}({\bf x}, t),
\label{u}
\end{equation}
where ${\bf u}({\bf x}, t)$ is the fluid velocity. Note here that the "particles" are $not$ the actual 
molecules that makes up the fluid, but an ensemble of infinitesimal fluid elements. Integrating 
Eq.~(\ref{klim}) using Eq.~(\ref{u}), we get
\begin{equation}
\partial_t {\bf u} + \nabla \cdot ({\bf u}{\bf u}) - {\bf a} = 0.
\label{NS}
\end{equation}
In the above, we have used the relation
\begin{equation}
{\bf u}{\bf u} = \int d{\bf v} {\bf v}{\bf v} f,
\label{mflux}
\end{equation}
reflecting the fact that a particle is moving exactly according to the fluid flow field.  Equivalently, it also indicates that the pdf $f({\bf x}, {\bf v}, t)$ of fluid elements has zero temperature (i.e., root-mean-square (rms) deviation of local fluid velocity), in contrast to the pdf of molecules corresponding to thermal temperature.

Equation (\ref{NS}) recovers the exact Navier-Stokes if the body-force term ${\bf a}$ is defined by
\begin{equation}
{\bf a} = - \nabla p + \nu \nabla^2 {\bf u},
\label{force}
\end{equation}
where $\nu$ is the molecular viscosity, and the pressure $p({\bf x}, t)$ is determined by the incompressibility constraint.

Now let us consider an ensemble averaged (or a coarse-grained) pdf, $F({\bf x}, {\bf v}, t) \equiv \langle f({\bf x}, {\bf v}, t)\rangle$.  Obviously, the normalization condition remains valid as
\begin{equation}
\int d{\bf v} F({\bf x}, {\bf v}, t) = 1.
\label{rho-bar}
\end{equation}
The averaged pdf gives the averaged velocity field, i.e., ${\bf U}({\bf x}, t) \equiv \langle {\bf u}({\bf x}, t)\rangle$, so we have
\begin{equation}
\int d{\bf v} {\bf v} F({\bf x}, {\bf v}, t) = {\bf U}({\bf x}, t)
\label{u-bar}.
\end{equation}

Taking the average over Eq.~(\ref{klim}), we obtain
\begin{equation}
\partial_t F + {\bf v}\cdot\nabla F + {\bar {\bf a}}\cdot\nabla_{\bf v} F = - \nabla_{\bf v}\cdot \langle \delta {\bf a} \delta f\rangle.
\label{boltz}
\end{equation}
In the above, ${\bf a} = {\langle \bf a \rangle} + \delta {\bf a}$ and $f = F + \delta f$, $F = \langle f \rangle$. Due to linearity and incompressibility, we have
\begin{equation}
{\langle {\bf a}\rangle} = - \nabla {\bar p} + \nu \nabla^2 {\bf U}
\label{force-bar}
\end{equation}
Because of the existence of the unknown term on the right-hand side, Eq.~(\ref{boltz}) for the averaged pdf is not closed.

Taking the average over the Navier-Stokes equation (i.e., Eq.~(\ref{NS})), we get the known Reynolds averaged equation
\begin{equation}
\partial_t {\bf U} + {\bf U}\cdot \nabla {\bf U} 
+ \nabla\cdot \sigma \; {-} {\langle {\bf a}\rangle}
= 0,
\label{RANS}
\end{equation}
where ${\langle {\bf a}\rangle}$ is given by (\ref{force-bar}). The additional quantity $\sigma$ is the so-called Reynolds stress tensor defined as
\begin{equation}
\sigma \equiv \langle \delta{\bf u}\delta{\bf u}\rangle.
\label{Re}
\end{equation}
From the averaged pdf, we have
\begin{eqnarray}
\int d{\bf v} {\bf v}{\bf v} F &=& \langle \int d{\bf v} {\bf v}{\bf v} f \rangle \nonumber \\
&=& \langle {\bf u}{\bf u} \rangle = {\bf U}{\bf U} + \sigma 
\label{mflux-bar}.
\end{eqnarray}
In the above, the basic definition ${\bf u} = {\bf U} + \delta {\bf u}$ and $\langle \delta {\bf u}\rangle = 0$.  Consequently, taking the trace of Eq.~(\ref{mflux-bar}), we get
\begin{equation}
\int d{\bf v} \frac {1} {2} {\bf v}^2 F 
= \frac {1} {2} \langle {\bf u}^2 \rangle
= \frac {1} {2} {\bf U}^2 + K, 
\label{K-equation}
\end{equation}
where $K \equiv \frac {1} {2} \langle (\delta {\bf u})^2\rangle$ is the turbulent kinetic energy.  

From Eqs.~(\ref{rho-bar}), (\ref{u-bar}), and (\ref{mflux-bar}) as well as Eq.~(\ref{K-equation}), one can recognize that $\sigma$ and $K$ have the following definitions based on the kinetic theory:
\begin{equation}
\int d{\bf v} ({\bf v} - {\bf U}) ({\bf v} - {\bf U})F = \sigma, 
\label{kin-def}
\end{equation}
and
\begin{equation}
\int d{\bf v} \frac {1} {2}({\bf v} - {\bf U})^2 F = K. 
\label{kin-def1}
\end{equation}
It is important to note from Eq.~(\ref{kin-def}) that the approach we developed here shows that the Reynolds stress tensor is fully determined by the averaged pdf, $F$. We emphasize the following non-trivial point. We have made no assumption so far that go beyond those usually attributed to the ensemble of turbulent fluctuations $\delta {\bf u}$; the entire information about this ensemble that is necessary for the full knowledge of turbulent Reynolds stress components, and thus for the full macroscopic description of turbulent flow, is contained in a single function $F({\bf x},{\bf v},t)$ that is determined by the kinetic theory. Further, Eq.~(\ref{kin-def1}) implies that the turbulent kinetic energy plays a role similar to the thermal energy in a regular kinetic theory for molecules. 

Integrating Eq.~(\ref{boltz}) by $\int d{\bf v}{\bf v}$, we have
\begin{equation}
\partial_t {\bf U} + {\bf U}\cdot \nabla {\bf U} 
+ \nabla\cdot \sigma \; {-} {\langle {\bf a}\rangle}
= - \int d{\bf v} {\bf v} \nabla_{\bf v}\cdot \langle \delta {\bf a} \delta f\rangle. 
\label{RANS-KT}
\end{equation}
Comparing Eqs.~(\ref{RANS}) and (\ref{RANS-KT}), the term on the right hand side of (\ref{RANS-KT}) should vanish. That is,
\[ 
- \int d{\bf v} {\bf v} \nabla_{\bf v}\cdot \langle \delta {\bf a} \delta f\rangle = 0. 
\]
This is also seen as a direct consequence of the fluctuation term on the right hand side of Eq.~(\ref{RANS-KT}). After integration by parts, we get
\[
- \int d{\bf v} {\bf v} \nabla_{\bf v} \langle \delta {\bf a} \delta f\rangle 
= \int d{\bf v} \langle \delta {\bf a} \delta f\rangle.
\]
Since ${\bf a}$ is not a function of ${\bf v}$, we have
\begin{eqnarray}
& & \int d{\bf v} \langle \delta {\bf a} \delta f\rangle 
= \langle \delta {\bf a} \int d{\bf v} \delta f\rangle
\nonumber \\ 
& & = \langle \delta {\bf a} \int d{\bf v} (f - F)\rangle = 0.
\end{eqnarray}
The result in the last step vanishes because $\int d{\bf v} (f - F) = 0$ from Eqs.~(\ref{rho}) and (\ref{rho-bar}).
Therefore, the additional term on the right-hand side of Eq.~(\ref{boltz}) is formally like a ``collision term" in the Boltzmann equation that conserves both mass and momentum:
\begin{eqnarray}
& & \int d{\bf v} \nabla_{\bf v}\cdot \langle \delta {\bf a} \delta f\rangle = 0 \nonumber \\
& & \int d{\bf v} {\bf v} \nabla_{\bf v}\cdot \langle \delta {\bf a} \delta f\rangle = 0.
\label{mm}
\end{eqnarray}
Consequently, we may rename this collision term in Eq.~(\ref{boltz}) as
\begin{equation}
C \equiv - \nabla_{\bf v}\cdot \langle \delta {\bf a} \delta f\rangle, 
\label{explicit-coll}
\end{equation}
but also point out that we do not know its explicit form. The central task in this turbulence representation is to find an appropriate closure for the collision term $C$.

\section{The energy equation}

Before investigating a possible form of $C$, we review the averaged energy equation here. Taking a dot product of ${\bf u}$ with Eqs.~(\ref{NS}) and (\ref{force}), we have
\begin{equation}
\partial_t (\frac {{\bf u}^2} {2}) + \nabla \cdot (\frac {{\bf u}^2} {2} {\bf u}) = - {\bf u}\cdot \nabla p + \nu {\bf u} \cdot \nabla^2 {\bf u}.
\label{KE}
\end{equation}
Likewise, taking a dot product of ${\bf U}$ with Eq.~(\ref{RANS}), we have,
\begin{equation}
\partial_t (\frac {{\bf U}^2} {2}) + \nabla \cdot (\frac {{\bf U}^2} {2} {\bf U}) + {\bf U} \cdot (\nabla \cdot \sigma ) = - {\bf U}\cdot \nabla {\bar p} + \nu {\bf U} \cdot \nabla^2 {\bf U}.
\label{U-KE}
\end{equation}
Substituting ${\bf u} = {\bf U} + \delta {\bf u}$ into Eq.~(\ref{KE}), taking the average, and subtracting Eq.~(\ref{U-KE}), after some algebra, we get
\begin{equation}
\partial_t K + {\bf U} \cdot \nabla K + \sigma : \nabla {\bf U}
+ \nabla \cdot ({\bf Q + \bf W})
= \nu \nabla^2 K - \epsilon.
\label{KE-bar}
\end{equation}
Here the dissipation term $\epsilon$ is given by
\begin{equation}
\epsilon \equiv \nu \langle (\nabla \delta {\bf u}) : (\nabla \delta {\bf u})\rangle,
\end{equation}
and the effective flux of turbulent kinetic energy due to turbulent fluctuations is given by
\begin{equation}
{\bf Q} \equiv \langle \delta {\bf u} \frac {(\delta {\bf u})^2} {2} \rangle.
\label{turb-flux}
\end{equation}
There is also a ``work"-related flux term $\nabla \cdot \bf W$, 
\begin{equation}
{\nabla \cdot \bf W} \equiv \langle {\delta{\bf u}\cdot 
\nabla \delta p}  \rangle,
\label{work}
\end{equation}
defined using the incompressibility condition ${ \nabla \cdot \delta \bf u} = 0.$

Equation (\ref{KE-bar}) governing the turbulent kinetic energy can also be derived straightforwardly from Eq.~(\ref{boltz}). It can be shown that the contribution from the collision term $C$ contains $\epsilon$.  Therefore, unlike the case of a regular molecular system, the turbulent kinetic energy is not conserved in its collision process among turbulent fluctuations, as expected from energy transfer from fluid fluctuations to thermal energy. Multiplying Eq.~(\ref{boltz}) by $\frac {1} {2} {\bf v}^2$ and performing $\int d{\bf v}$, we get
%
\begin{align}
\partial_t \int d{\bf v} \frac {1} {2} {\bf v}^2F
+ \nabla \cdot \int d{\bf v} {\bf v} \frac {1} {2} {\bf v}^2F
+ \int d{\bf v} \frac {1} {2} {\bf v}^2 {\bar {\bf a}}\cdot\nabla_{\bf v} F \nonumber
\\
= \int d{\bf v} \frac {1} {2} {\bf v}^2 C, 
\label{TKE-KT}
\end{align}
where $C$ is given by Eq.~(\ref{explicit-coll}).  According to Eq.~(\ref{K-equation}), 
\[ 
\int d{\bf v} \frac {1} {2} {\bf v}^2F = \frac {1} {2} {\bf U}^2 + K,
\]
where $K$ is defined in Eq.~(\ref{kin-def1}).  Integrating by parts, it can be shown that
\[
\int d{\bf v} \frac {1} {2} {\bf v}^2 {\langle {\bf a}\rangle}\cdot\nabla_{\bf v} F
= {\bf U} \cdot {\langle {\bf a}\rangle}. 
\]
With some straightforward algebra, one can show that
\[
\nabla \cdot \int d{\bf v} {\bf v} \frac {1} {2} {\bf v}^2F
= \nabla \cdot \{ {\bf U} [\frac {1} {2} {\bf U}^2 + K] 
+ {\bf U} \cdot \sigma 
+ {\bf Q} \},
\]
where $\sigma$ is defined in Eq.~(\ref{kin-def}); the flux $\bf Q$, defined by (\ref{turb-flux}), can be rewritten as
\begin{equation}
{\bf Q} \equiv \langle \delta {\bf u} \frac {(\delta {\bf u})^2} {2} \rangle  = \int d{\bf v} \frac {1} {2} ({\bf v} - {\bf U}) ({\bf v} - {\bf U})^2 F.
\label{kinetic defined Q}
\end{equation}
Taking all these into Eq.~(\ref{TKE-KT}), we obtain
\begin{align}
\partial_t [\frac {1} {2} {\bf U}^2 + K]
+ \nabla \cdot \{ {\bf U} [\frac {1} {2} {\bf U}^2 + K] 
+ {\bf U} \cdot \sigma 
+ {\bf Q}\}
+ {\bf U} \cdot {\bar {\bf a}} \nonumber \\
= \int d{\bf v} \frac {1} {2} {\bf v}^2 C.
\label{TKE-KT-int}
\end{align}
Subtracting Eq.~(\ref{U-KE}) from (\ref{TKE-KT-int}) and using the definition of average force (\ref{force-bar}), we obtain the following equation for the turbulent kinetic energy evolution from this kinetic theory formulation:
\begin{equation}
\partial_t K
+ {\bf U} \cdot \nabla K 
+ \sigma : \nabla {\bf U}
+ \nabla \cdot {\bf Q} 
= \int d{\bf v} \frac {1} {2} {\bf v}^2 C.
\label{TKE-KT-f}
\end{equation}
By recalling the definition (\ref{explicit-coll}), 
the right side of Eq.~(\ref{TKE-KT-f}) can be recast as
\begin{equation}
\int d{\bf v} \frac {1} {2} {\bf v}^2 C = - \nabla\cdot { \bf W}  +
\nu \nabla^2 K - \epsilon,
\label{epsilon-dissNew}
\end{equation}
to recover Eq.~(\ref{KE-bar}) for incompressible flows.

%
%

\section{The collision term}

It is difficult to derive an explicit form for the collision term $C$ in Eq.~(\ref{boltz}). In our previous work \cite{chen2}, we used an experimental observation that the single-particle pdf $F^{eq}$ in a homogeneous turbulent flow exhibits a local Gaussian form
\begin{equation}
F^{eq} = W\;exp[- 3({\bf v} - {\bf U})^2/2K],
\label{pdf-eq}
\end{equation}
where $W$ is the normalization factor for $\int d{\bf v} F^{eq} = 1$. See also \cite{sree1,sree2}. Let us now present some further semi-theoretical arguments for this form of the equilibrium distribution function, and argue that it is the asymptotic limit of the collision process $C$. Suppose we start evolving the distribution function $F$ according to Eq.~(\ref{boltz}) towards its local equilibrium $F^{eq}$. Based on the expression for the collision integral in Eq.~(\ref{explicit-coll}), the ensemble average could be reinterpreted as
\begin{equation}
C = - \langle \delta {\bf a} \cdot \nabla_{\bf v}\delta f\rangle
= - \frac {1} {N} \sum_n^N \delta {\bf a}^{(n)} \cdot \nabla_{\bf v}\delta f^{(n)}.
\label{ensemble}
\end{equation}
Here the ensemble is explicitly written as the average of $N$ realizations of deviations from the mean, for $n = 1, \ldots , N \rightarrow \infty$.  Therefore Eq.~(\ref{ensemble}) defines $C$ as the summation of $N$ small increments to the distribution function as it evolves towards equilibrium, as defined by Eq.~(\ref{boltz}). By an informal appeal to the central limit theorem, the argument can be made that $F$ should converge, in the most probable sense, to a Gaussian form, since it is the average sum of infinitely many uncorrelated processes. We realize that, while realizations are uncorrelated, their deviations from the mean are not; we are thus fully self-critical with respect to this argument. If, however, one accepts that these processes are uncorrelated to a leading order, then the only Gaussian that can be a candidate for the equilibrium is that given by Eq.~(\ref{pdf-eq}), since it satisfies the conservation of mass and momentum, as well as the energy constraint based on (\ref{rho-bar}), (\ref{u-bar}) and (\ref{K-equation}): 
\begin{eqnarray}
& & \int d{\bf v} F^{eq} = \int d{\bf v} F = 1 \nonumber \\
& & \int d{\bf v} {\bf v} F^{eq} = \int d{\bf v} {\bf v} F = {\bf U}
\nonumber \\
& & \int d{\bf v} \frac {1} {2} {\bf v}^2 F^{eq}  
= \frac {1} {2} {\bf U}^2 + K^{\prime}.
\label{3constraints}
\end{eqnarray}
These constraints together with a Gaussian distribution completely determine $F^{eq}$ given by Eq.~(\ref{pdf-eq}). Here $K^{\prime}$ is the value of turbulent kinetic energy in equilibrium. Therefore, the collision process $C$ drives the distribution towards an equilibrium distribution which is exactly defined by Eq.~(\ref{pdf-eq}).  We note that this semi-theoretical argument does not quantify the rate of approach to equilibrium; that is, we do not know the time it takes for the collision process $C$ to drive $F$ to $F^{eq}$.

Based on the argument above, we could assume that the collision term can be described as a relaxation process towards the local Gaussian equilibrium with an unspecified relaxation time $\tau$, namely a BGK form
\begin{equation}
C = - \frac {F - F^{eq}} {\tau}.
\label{BGK}
\end{equation}
Equation (\ref{boltz}) together with (\ref{BGK}) gives a closed kinetic Boltzmann-BGK representation for the averaged fluid motion with an unspecified relaxation time $\tau$. Obviously, $\tau = \tau ({\bf x}, t)$, a function of local and nonlocal properties and structures of turbulence. Understanding and the determination of this fundamental time scale is work for the future, but a few comments are in order.

First, $\tau$ must have similar characteristics as $K/\epsilon$ but it must be kept in mind that the deviation of $F$ from the equilibrium value, $F^{eq}$, occurs when there exists a shear in the mean flow, as dictated by Eq.~(\ref{boltz}). This is in agreement with the experimental observation of the Gaussian single-point pdf $F^{eq}$ (Eq.~(\ref{pdf-eq})) in homogeneous turbulence. Second, one can in principle write down a dynamic equation for $\tau$ as done in a different context in Ref.~\cite{prab}, where a wake behind a twin-airfoil, driven away from equilibrium by an impulsive pressure gradient, was allowed to relax back to equilibrium.

As discussed above, the collision operator $C$ obeys the conservation of mass and momentum in (\ref{mm}) as 
\begin{eqnarray}
& & \int d{\bf v} [F - F^{eq}] = 0 \nonumber \\
& & \int d{\bf v} {\bf v} [F - F^{eq}] = 0.
\label{coll-mm}
\end{eqnarray}
In contrast, $C$ does not conserve the turbulent kinetic energy $K$. As shown above, the turbulent kinetic energy should change, in the sense of a global space average,  according to
\begin{equation}
\int d{\bf v} \frac {({\bf v} - {\bf U})^2} {2} C = - \epsilon
\label{coll-e}.
\end{equation}
Consequently, the $K$ value appearing in $F^{eq}$ (Eq.~(\ref{pdf-eq})) should be slightly smaller than the $K$ value derived using $F$ according to Eq.~(\ref{K-equation}) or Eq.~(\ref{kin-def1}). If we assume the BGK form for the collision term, then the difference between the two $K$ values is simply $\Delta K = \tau \epsilon$.

\section{Kinetic representation of the average turbulence dynamics}

Equations (\ref{rho-bar}, \ref{u-bar}), (\ref{kin-def1}), (\ref{boltz}, \ref{force-bar}), (\ref{explicit-coll}), (\ref{pdf-eq}), (\ref{3constraints}, \ref{BGK}), and (\ref{coll-e}) fully define a kinetic theoretic representation of the averaged turbulent flow dynamics in the form of Boltzmann-BGK with only one unspecified collision relaxation time, $\tau$. Solving the kinetic system one obtains the averaged pdf $F$, and the latter fully determines fundamental quantities such as the Reynolds stress tensor $\sigma$ according to Eq.~(\ref{kin-def}). Note that for homogeneous flow the derivatives in the kinetic equations (\ref{boltz}) and (33) disappear so that the
solution corresponding to the BGK collision integral (\ref{coll-e}) is the Gaussian $F^{eq}$ (\ref{pdf-eq}), which is consistent with experimental observations for single point pdf. Therefore, from this perspective, the non-Gaussian statistics are due to spatial inhomogeneities of the averaged flow, generating non-equilibrium properties. 

It is straightforward to derive that Eqs.~(\ref{RANS}) and (\ref{KE-bar}) are the results of the appropriate moments of the Boltzmann-BGK model. The resulting Reynolds stress term has an explicitly expanded form in terms of turbulent hydrodynamic quantities via the Chapman-Enskog expansion of a finite mean-free path \cite{chen}; its leading order (i.e., Navier-Stokes) expression (in the small mean-free path limit) is the usual eddy-viscosity form
\begin{equation} 
\sigma_{ij} = \frac {2} {3} K \delta_{ij} - \nu_{eddy} {\bar S}_{ij},
\label{eddy-visc}
\end{equation}
where $\nu_{eddy} = 2\tau K/3$. Here $\delta_{ij}$ is the Kronecker delta function, and ${\bar S}_{ij}$ is the averaged strain rate tensor
\[ 
{\bar S}_{ij} = \frac {1} {2} (\frac {\partial U_i} {\partial x_j} + \frac {\partial U_j} {\partial x_i} ).
\]
The trace of ${\bar S}_{ij}$ vanishes due to incompressibility. The next order contributions to $\sigma$ include both time and space derivatives and nonlinear functions of ${\bar S}_{ij}$ as well as the anti-symmetric tensor ${\bar \Omega}_{ij}$ \cite{chen}, as given below:
\begin{eqnarray}
\sigma^{(2)}_{ij} = &-& 2\nu_{eddy} \frac {D} {Dt} [\tau {\bar S}_{ij}]
\nonumber \\
&-& 6 \frac {\nu_{eddy}^2} {K} [{\bar S}_{ik}{\bar S}_{kj} 
- \frac {1} {3} {\bar S}_{kl} {\bar S}_{kl}] \nonumber \\
&+& 3 \frac {\nu_{eddy}^2} {K} [{\bar S}_{ik} {\bar \Omega}_{kj} 
+ {\bar S}_{jk} {\bar \Omega}_{ki}].
\label{sigma-2}
\end{eqnarray} 
Here, $D/Dt \equiv \partial_t + {\bf U}\cdot \nabla$, and 
\[ 
{\bar \Omega}_{ij} = \frac {1} {2} (\frac {\partial U_i} {\partial x_j} - \frac {\partial U_j} {\partial x_i} )
\]
is the vorticity tensor. The summation convention for repeated indices is adopted in Eq.~(\ref{sigma-2}). It is interesting to mention that by associating $\tau$ to a known turbulent eddy viscosity model, the coefficients in the next order (i.e., (\ref{eddy-visc}) and (\ref{sigma-2})) are in quantitative agreement with some representative nonlinear turbulence models that were constructed or proposed from various other means \cite{chen}.  
Using the same analysis, one can also expect, to the leading order,
\begin{equation} 
{\bf Q} = - \eta_{eddy} \nabla K
\label{eddy-visc-K},
\end{equation}
with $\eta_{eddy} \sim \nu_{eddy}$.

In short, we have theoretically constructed a kinetic theory representation for the averaged turbulent flow dynamics. However, the equation is not closed due to the unspecified time $\tau$. The latter is central for formulating any specific kinetic theory based turbulent models. This requires an understanding of the structures and interactions (collisions) of turbulent fluctuations in terms of $\langle \delta {\bf a} \cdot \nabla_{\bf v} \delta f \rangle$. 

\section{Discussion}  
\subsection{Past closure attempts}
Closure models for the Reynolds stress tensor with an eddy-viscous form as in (\ref{eddy-visc}) (and (\ref{eddy-visc-K})) has been a long standing foundation for mainstream turbulence models. The underlying concept has been the heuristic analogy between turbulent eddy fluctuations and molecular motions. For the mean motion of turbulent flows, this analogy results in Navier-Stokes equation with an added eddy viscosity. However, this procedure is questionable because of the lack of scale separation between the mean flow and the so-called turbulent eddies. (Separate studies \cite{chen} have also demonstrated weak correlation between the Reynolds stress tensor $\sigma$ and the mean turbulent rate of strain ${\bar {\bf S}}$ tensor.)  Extensions beyond the eddy viscosity term, which include higher order non-Newtonian effects with the framework of a modified Navier-Stokes equation \cite{spez,yosh,rubi,yakh}, have offered useful insights but limited success.  

In the past, kinetic theory representation based on the Boltzmann equation has also been attempted \cite{dego,chen,chen2}. However, these attempts are imposed externally with the intent to generate the {\it a priori} known averaged turbulent hydrodynamic equations, although the former contains more information beyond the viscous order \cite{chen}. There is a question as to why such a kinetic equation should be justified other than producing the already expected hydrodynamic equations. There is also the question of how such a kinetic representation related to the first principles of turbulence. On the other hand, one main issue from averaging the regular Boltzmann equation is that the resulting averaged equilibrium distribution contains the temperature in the same place as the turbulent kinetic energy. Hence, the thermal molecular effects cannot be removed by successive averaging procedure that can pick out the tiny hydrodynamic velocity from a microscopic velocity of molecules. Thus, the most straightforward approach of performing averages of the regular Boltzmann equation over molecular motions \cite{chen3}) has been without commensurate gain (see also \cite{Girimaji}). 

\subsection{Present work}

We have shown here that the starting point for deriving a desired kinetic equation for averaged dynamics of turbulence can be a Klimontovich-type self-consistent body-force-based equation. This equation describes the motion of particles (i.e., fluid elements of infinitesimal size) corresponding exactly to the (unaveraged) Navier-Stokes hydrodynamics. Averaging the Klimontovich equation results in a kinetic equation with a collision term representing the effect of turbulent fluctuations on mean dynamics. Our present kinetic theoretic representation does not suffer from the problem of scale separation. Besides including the eddy viscosity as well as the next order effects, effects from turbulent fluctuations of all orders of a scale expansion are also naturally included.

An attractive feature of this kinetic-based approach is that a single distribution function $F({\bf x}, {\bf v}, t) \equiv \langle f({\bf x}, {\bf v}, t)\rangle$, governed by (\ref{boltz}), contains the entire information needed for the (turbulent ensemble) averaging of all properties, as long as they are based on single point pdfs. Indeed, in addition to second order moments (\ref{kin-def}, \ref{kin-def1}), the knowledge of $F$ provides full information about the turbulent energy diffusion flux (\ref{kinetic defined Q}) which is a third order quantity---and, indeed, about any arbitrary N-th order moment of the single point pdf of turbulent fluctuations.  In particular, the non-Gaussian properties of the pdf such as flatness and skewness are all contained in the kinetic-based distribution function $F$. Note also that the all these conclusions are obtained under some of the weakest assumptions possible. Indeed nothing stronger than the mere existence of ensemble average (such that $\langle \delta \bf u \rangle = 0$) is required to exactly derive expressions such as (\ref{kin-def}), (\ref{kin-def1}), and (\ref{kinetic defined Q}); in particular no assumptions about the specific form of the equilibrium distribution function and/or collision integral are invoked. This self-consistent approach can be applied for the kinetic-theory reformulation of equations other than Navier-Stokes (e.g., the Burgers equation) by an appropriate choice of self consistently generated body force ${\bf a}({\bf x},t)$ (Eq.~(\ref{force})).

In this formulation, the turbulent kinetic energy $K$, which measures the square of the mean turbulent fluctuation, plays a role like the thermal energy or temperature in the regular Boltzmann equation for motion of molecules. It is interesting to note one important difference between a turbulent kinetic energy based temperature $T_{turb}$ ($= 2K/3$) and a conventionally known thermal temperature $T$: Unlike a regular low Mach number flow with $\sqrt{T} >> |{\bf u}|$, the turbulent kinetic energy based temperature is often less than the mean flow velocity, $\sqrt{T_{turb}} < |{\bf U}|$. Therefore, an averaged turbulent flow bears some character of a supersonic flow, even though it is still incompressible because of the pressure term from the incompressibility constraint.  

We have shown that the averaged kinetic equation generates the same form of the governing equations for the mean turbulent velocity ${\bf U}$ and the turbulent kinetic energy $K$, as that from direct averaging of the Navier-Stokes equation. Furthermore, interactions of turbulent fluctuations play a role like a molecular collision process that drives the mean pdf towards a Maxwell-Boltzmann (Gaussian) equilibrium distribution. The single relaxation time scale determines not only the eddy viscosity, but also the turbulent transport coefficients of all orders in the expansion of turbulent flow scales. The collision process obeys the conservation of mass and momentum. On the other hand, it generates a loss of flow energy that directly corresponds to the turbulent dissipation $\epsilon$. The present analysis justifies a relaxation process to equilibrium. The full kinetic equation for describing the dynamics of a mean turbulent flow is closed when the relaxation time $\tau$, the dissipation $\epsilon$, as well as the flux $\bf W$ (cf.  Eqs.~(\ref{epsilon-dissNew}, \ref{work})), are specified. However, whether all turbulent moments relax with the same relaxation time $\tau$ (i.e., BGK model) is a model assumption, although there is some supporting evidence for this view \cite{yakh}.

\subsection{The collision term and the relaxation effects}

An estimate, based on a typical conventional turbulence model, indicates that the effective mean-free path, $L_{mfp}$, or the collision relaxation time $\tau$ are, respectively, about one tenth of the length and time scale of the energy containing eddies, namely $L_{mfp} \sim C_{\mu}K^{3/2}/\epsilon$ and $\tau \sim C_{\mu} K/\epsilon$. $C_{\mu} \sim 0.1$. Therefore, a mean turbulent flow is more resembling of a flow with finite Knudsen number, $Kn \sim 0.1$. This suggests that the representation of the mean turbulent flow dynamics by the Navier-Stokes equation with an eddy-viscosity is not sufficient, in general. Instead, contributions from the higher order terms (in the small Knudsen number expansion) become non-negligible, and are responsible for non-Newtonian fluid dynamic behavior beyond the Navier-Stokes representation. Indeed, averaged turbulence exhibits the known phenomena of finite Knudsen number and non-Newtonian character, such as the slip velocity near wall \cite{chen2}, the secondary flow structures in a square duct and the known retarded rapid distortion behavior (cf. \cite{spez,rubi,yakh}). A kinetic theory based representation for turbulence naturally includes finite Knudsen effects of all orders, and thus all the non-Newtonian phenomena mentioned above are automatically captured. 
 
Lastly, we wish to make the following observation. Microscopic motion of molecules is described by a Boltzmann-type kinetic equation. The averaged motion of the molecules is described by the Navier-Stokes hydrodynamic equation. Yet, when further averaged, a suitable kinetic formulation comes back as a desirable description---completing a virtuous circle. 

\vspace{0.1in}

{\bf Acknowledgment:} We acknowledge the influence on this work of extended interactions with our late friend, Steven A. Orszag.

\end{document}